\documentstyle[aps,prl,graphicx]{revtex}

\input psfig.sty
\draft
\begin{document}
\twocolumn 
\wideabs{  
\title{Observation of Long-lived Vortex Aggregates in Rapidly Rotating Bose-Einstein Condensates}
\author{P. Engels, I. Coddington, P.~C. Haljan, V. Schweikhard, and E.~A. Cornell\cite{qpdNIST}}
\address{JILA, National Institute of Standards and Technology and University of Colorado,
and Department of Physics, University of Colorado, Boulder,
Colorado 80309-0440}
\date{\today}

\maketitle

\begin{abstract}
We study the formation of large vortex aggregates in a rapidly
rotating dilute-gas Bose-Einstein condensate. When we remove
atoms from the rotating condensate with a tightly focused,
resonant laser, the density can be locally suppressed, while
fast circulation of a ring-shaped superflow around the area of
suppressed density is maintained. Thus a giant vortex core
comprising 7 to 60 phase singularities is formed. The giant core
is only metastable, and it will refill with distinguishable
single vortices after many rotation cycles. The surprisingly
long lifetime of the core can be attributed to the influence of
strong Coriolis forces in the condensate. In addition we have
been able to follow the precession of off-center giant vortices
for more than 20 cycles.

\end{abstract}

} 
\par

The phenomenon of quantized vortex formation is a unifying
feature present in many quantum mechanical systems. Vorticity is
intimately connected to superfluidity and thus  is a premier
means for fundamental  and comparative studies of different
``super" systems. In dilute gas Bose-Einstein condensates,
vortex experiments have ranged from the study of individual or
few vortices
\cite{firstvortex,Vortexprecession,Madison2000a,Hodby2001a} and vortex rings \cite{vortexrings,HauScience} to
the first creation of vortex lattices
\cite{Madison2001a} and the study of systems containing large
amounts of vorticity \cite{Abo-Shaeer2001a,Jilanucleation}.

Using a refinement of our experimental technique, we have now
been able to generate a giant vortex. We will use the term
``giant vortex" to denote a region, containing multiple units of
vorticity, in which the density is completely suppressed such
that the individual vortices are no longer discernible. We note
that some authors reserve this term for higher order phase
singularities.
 The possible existence of stable giant vortices under certain conditions like supersonic
flow in potentials that have strong quartic terms has been
theoretically predicted
(Ref.~\cite{GiantvortexLundh,GiantvortexUeda,GiantvortexBaym},
see also Ref.~\cite{QuarticpotentialFetter}), and in
Ref.~\cite{GiantvortexSalomaa} it is shown that a pinning
potential can lead to stable multiquantum vortices. In our case,
however, the giant vortex formation arises as a dynamic effect.
Nevertheless the lifetime of our giant vortices can extend over
many seconds, which we attribute to a stabilization of density
features in a rapidly rotating condensate due to strong Coriolis
forces. The influence of Coriolis forces can also induce
oscillations of the giant vortex core size in the early stages
of its evolution.

\par
\begin{figure}
\begin{center}
\psfig{figure=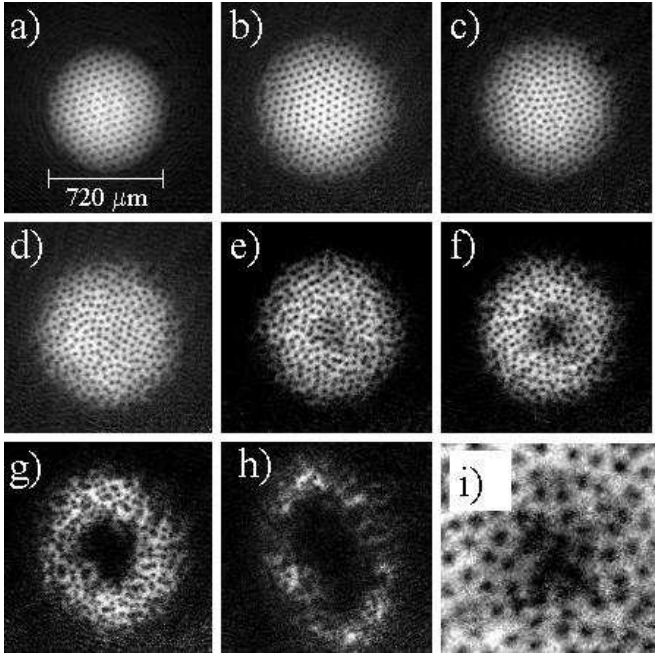,width=1\linewidth,clip=}
\end{center}
\caption {Different stages of giant vortex formation process. (a) Starting point: BEC after evaporative spin-up.
(b)-(h) Laser shone onto BEC for (b) 14~s, (c) 15~s, (d) 20~s,
(e) 22~s, (f) 23~s, (g) 40~s, (h) 70~s. Pictures are taken after
5.7-fold expansion of the BEC. Laser power is 8~fW. (i)
Zoomed-in core region of (f).}
\label{giantexpansions}
\end{figure}
\begin{figure}
\begin{center}
\psfig{figure=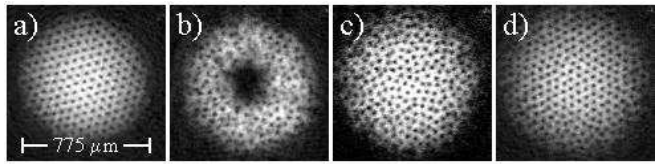,width=1\linewidth,clip=}
\end{center}
\caption {Lattice reforming after giant vortex formation. (a) BEC after evaporative spin-up.
(b) Effect of a 60~fW, 2.5~s laser pulse. (c),(d) Same as (b),
but additional 10~s (c) and 20~s (d) in-trap evolution time
after end of laser pulse. Images taken after a sixfold expansion
of the BEC.}
\label{latticeholelattice}
\end{figure}

Our experimental starting conditions are similar to the ones
described in our previous papers
\cite{Jilanucleation,Jilastripes}. We start with a cloud of
uncondensed $^{87}$Rb atoms in the $|F=1,m_{F}=-1\rangle$ state,
held in a magnetic TOP trap and precooled to a temperature
slightly above the critical temperature for Bose-Einstein
condensation. By elliptically deforming the magnetic trap in the
horizontal xy plane and suddenly changing the angle of the
deformation, we impart angular momentum around the vertical
z-axis to the cloud. We then restore axial symmetry and perform
a quasi one-dimensional evaporation in a trap with frequencies
$\{\omega_{\rho},\omega_{z}\}=2\pi\{8.3,5.4\}$Hz. This
evaporation preferentially removes atoms close to the axis of
rotation (z-axis), thus spinning up the cloud. At the end of
this step, we routinely achieve fast rotating condensates
consisting of $3\times10^{6}$ atoms with a rotation rate of 95\%
of the radial trap frequency, with no detectable thermal cloud.
This rotation rate is deduced from the aspect ratio, which
changes from the value of 1.53 for a static cloud to 0.48
because of the centrifugal forces created by the rotation. These
condensates typically contain 180 or more vortices as seen in
Fig.~\ref{giantexpansions}(a) or
Fig.~\ref{latticeholelattice}(a). After careful optimization of
the trap roundness and in the presence of a quasi 1-D rf-shield
we are able to observe the BEC rotation for times exceeding
5~minutes. While we do lose atoms from the condensate over this
time scale, the rotation rate remains at its initial value.

The evaporative spin-up mechanism is limited because eventually
all the uncondensed atoms have been removed or condensed during
the quasi one-dimensional evaporation. However, the angular
momentum per particle in the condensate can be further enhanced
by selectively removing atoms with low angular momenta by a
resonant, focused laser beam sent through the condensate on the
axis of rotation.  Since the rotating BEC typically has a
Thomas-Fermi radius of $66\mu m$, the width of the laser beam of
about $16 \mu m$, stated here as the full width at half maximum
(FWHM) of the Gaussian intensity profile, is small enough to
provide sufficient selectivity. The frequency of the laser is
tuned to the $F''=1
\rightarrow F' = 0$ transition of the D2 line, and the recoil
from spontaneously scattered photons blasts atoms out of the condensate.

\begin{figure}
\begin{center}
\psfig{figure=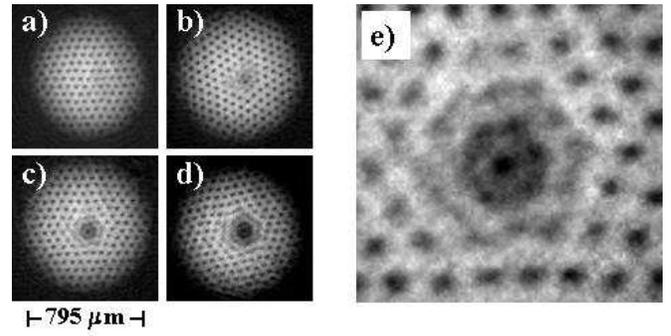,width=1\linewidth,clip=}
\end{center}
\caption {Core developing after a 5~ms short, 2.5~pW laser pulse.
 In-trap evolution time after end of pulse is (a) 0.5~ms, (b)
10.5~ms, (c) 20.5~ms, (d) 30.5~ms. Images taken after sixfold
expansion of BEC. (e) Zoomed-in core region of (c).}
\label{coreopening}
\end{figure}

\begin{figure}
\begin{center}
\psfig{figure=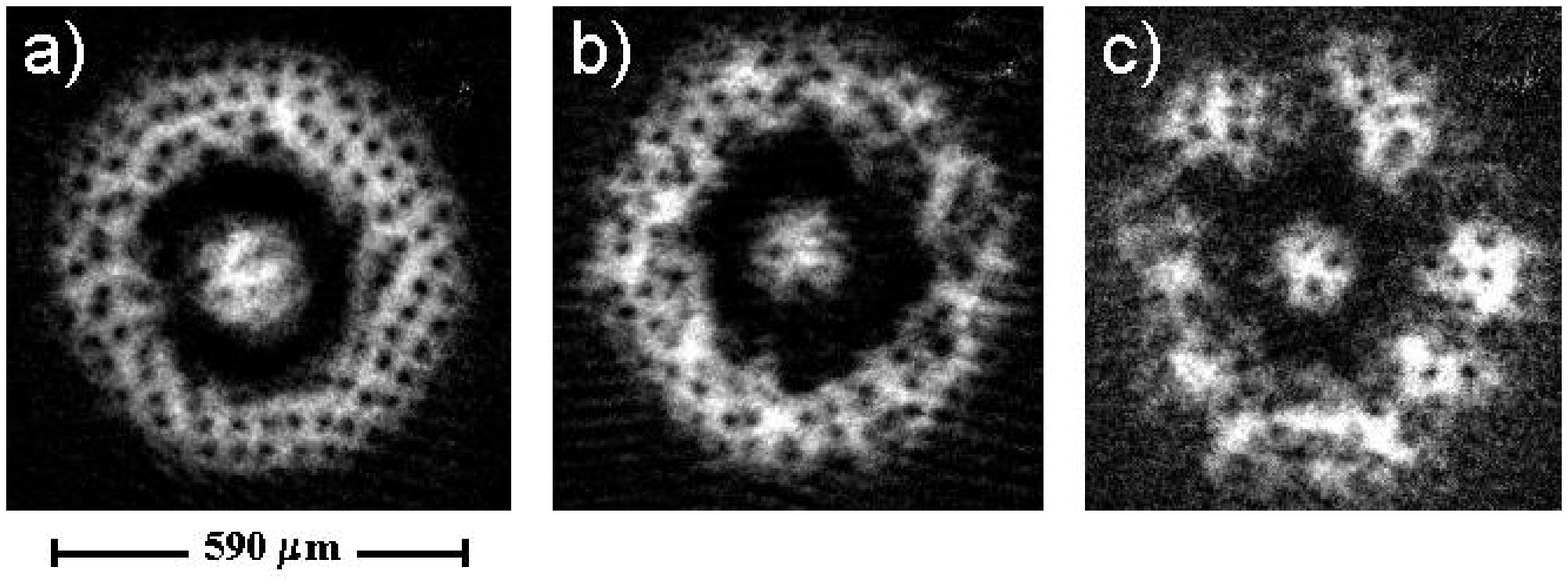,width=1\linewidth,clip=}
\end{center}
\caption {Ring cut out of a BEC by a 125~ms long, off-centered laser pulse. Expansion image taken
(a) directly after end of pulse (b) after an additional in-trap
evolution time of 200 ms after end of pulse and (c) evolution
time 2~s. }
\label{cutring}
\end{figure}

Using this new technique we are able to substantially increase
the vorticity in the BEC and can now routinely obtain
condensates containing more than 250 vortices
[Fig.~\ref{giantexpansions}(b)]. Since these  vortex cores are
too small to be imaged when the BEC is held in trap, the images
in Figs.~\ref{giantexpansions},
\ref{latticeholelattice},
\ref{coreopening}, and \ref{cutring} are taken after having let the
BEC expand. For this, a rapid adiabatic passage technique is
employed to change the hyperfine state of the atoms from the
original trapped $|F=1, m_{F} = -1\rangle$ state to the
antitrapped $|F=2, m_{F}=-1\rangle$ state. Subsequently the
minimum of the magnetic trapping field is rapidly moved below
the position of the atoms so that the atoms are supported
against gravity while at the same time they are radially
expelled \cite{Heather}. After a chosen expansion time the
magnetic fields are switched off and the atoms are imaged along
the original axis of rotation. This technique allows us to
expand the condensate to an adjustable diameter that can exceed
1.4~mm
\cite{MacDonald2002}.

\par
If the removal of central atoms is driven strongly enough, a
giant vortex appears. The core is surrounded by a ringshaped
superflow. The circulation of this superflow is given by summing
up all the ``missing" phase singularities in the core and can
assume very large values. Using the fact that for large amounts
of circulation the rotation of the velocity field can be
obtained classically, we determine that in some cases the BEC
supports a ring-shaped superflow with a circulation of up to 60
quanta around the core. We can easily control the amount of this
circulation by changing the intensity or the duration of the
laser beam that removes atoms along the axis of rotation.

In our experiment the formation of the giant vortex comes about
in a sequence of very distinct stages as shown in
Fig.~\ref{giantexpansions}. For this expansion image sequence a
rapidly rotating BEC is first formed by our evaporative spin-up
technique. Then the atom-removal laser is applied with a fixed
power of 8~fW for a variable amount of time, followed by a 10~ms
in-trap evolution time and a 45~ms expansion in the antitrapping
configuration described above. During the atom removal an
rf-shield is left on to constantly remove uncondensed atoms that
tend to decelerate the condensate rotation.
Figure~\ref{giantexpansions}(a) shows the result of only the
evaporative spin-up. This particular condensate contains 180
vortices and has a Thomas-Fermi radius of $63.5\,\mu m$ when
held in the trap. When the atom removal laser is applied for
14~s as in Fig.~\ref{giantexpansions}(b), the number of vortices
is increased to 250 and the Thomas-Fermi radius to $71\,\mu m$.
The rotation rate, determined from sideview aspect-ratio images,
has increased from $0.94\,\omega_{\rho}$  to
$0.97\,\omega_{\rho}$.  After atom removal times of 15 to 20~s,
the vortex lattice becomes disordered
[Fig.~\ref{giantexpansions}(c,d)], and the giant vortex core
starts to develop in the center
[Fig.~\ref{giantexpansions}(e,f)]. An enlarged view of the core
region in this early stage of giant vortex core formation is
seen in Fig.~\ref{giantexpansions}(i), which nicely demonstrates
how the individual vortices in the center consolidate. For the
larger removal times shown in Figs.~\ref{giantexpansions}(g) and
(h), a clear elliptical deformation is observed. This
deformation is due to residual static asymmetries of the
trapping potential that can nearly resonantly drive quadrupolar
surface modes when the rotation frequency of the condensate
approaches the radial trapping frequency. We have verified this
explanation by deliberately changing the roundness of the
trapping potential and observing corresponding changes in the
ellipticity and in the direction of the ellipse. If the trap
roundness is not optimized, the ring-shaped superflow has a
tendency to fragment into typically 5 to 8 blobs that continue
to rotate around the core.

\par
If the atom removal laser is applied only for a limited amount
of time, the giant vortex will fill in and the vortices refreeze
\cite{KetterleCrystallization} into a well ordered lattice again
after a sufficiently long evolution time. An example is shown in
Fig.~\ref{latticeholelattice}. Here, the starting point is the
condensate in Fig.~\ref{latticeholelattice}(a), containing 190
vortices and rotating at $0.95\,\omega_{\rho}$. A pulse from the
atom removal laser creates the core seen in
Fig.~\ref{latticeholelattice}(b). Crudely assuming that all
atoms, and only those atoms, originally within a cylindrical
volume of one-third of the original cloud radius are removed, we
can perform an integral over the Thomas-Fermi density profile
with a rigid-body-rotation velocity distribution, and find that
23\% of the atoms have been removed but only 4\% of the angular
momentum. Since in equilibrium the total number of vortices
should be linear in the average angular momentum per atom
\cite{Feder}, we would expect that the original 190-vortex cloud
should reequilibrate to form a cloud with 236 vortices, in rough
agreement with the value of 260 vortices observed in the
reequilibrated cloud shown in Fig.~\ref{latticeholelattice}(d).

\par
In our case the formation of the giant vortex is not connected
to a repulsive conservative optical potential produced by the
central laser beam, but only due to the removal of atoms from
the axis of rotation by spontaneous photon scattering. To prove
this, we have varied the laser frequency over the range from
-6~MHz to +6~MHz around the $F''=1
\rightarrow F' = 0$ resonance and in all cases have been able to
generate a giant vortex.

Since theoretical studies predict the appearance of a giant
vortex if the trapping potential has a sufficiently strong
admixture of quartic terms, it is important to quantify the
strength of the quartic terms in our case. For this we have
calculated the term of the TOP trap potential that depends on
the radial position as $r^{4}$. At the outer edge of the
condensate this quartic term produces a correction to the
potential of less than $10^{-3}$, compared to the effective
$r^{2}$ term, even though the latter is much weakened by the
centrifugal force. Higher order corrections contribute even
less. Empirically, the refilling and refreezing we observe
(Fig.~\ref{latticeholelattice}) indicates that the giant vortex
is not an equilibrium configuration but has instead only
dynamical stability.

We suggest an intuitive, classical picture describing the
formation as a dynamical effect: the removal of atoms from the
center of the condensate produces a pressure gradient due to
mean field energy that tries to drive atoms from the outer
regions into the center, so as to close the hole. Due to
Coriolis forces, however, atoms moving radially towards the
center are deflected and assume a fast azimuthal motion around
the core rather than filling the core, thus creating the giant
vortex.

Some exotic features of the early stages of core formation are
revealed when we apply only a very short, weak atom removal
pulse and observe the subsequent evolution as in
Fig.~\ref{coreopening}. Here, the atom removal laser has a power
of 2.5~pW and a fixed pulse length of 5~ms. Its FWHM of $16\,
\mu m$ is approximately twice the lattice spacing, $7 \mu m$.
By varying the delay time between the laser pulse and the
expansion, we see (Fig.~\ref{coreopening}) that the core
formation clearly lags behind the laser pulse.
Fig.~\ref{coreopening}(e) shows a zoomed-in view of the core
region of Figure~\ref{coreopening}(c). It is very interesting to
observe that for these short, weak pulses the density depression
appears to develop in discrete steps, and the step boundaries
follow the hexagonal lattice pattern.

\begin{figure}
\begin{center}
\psfig{figure=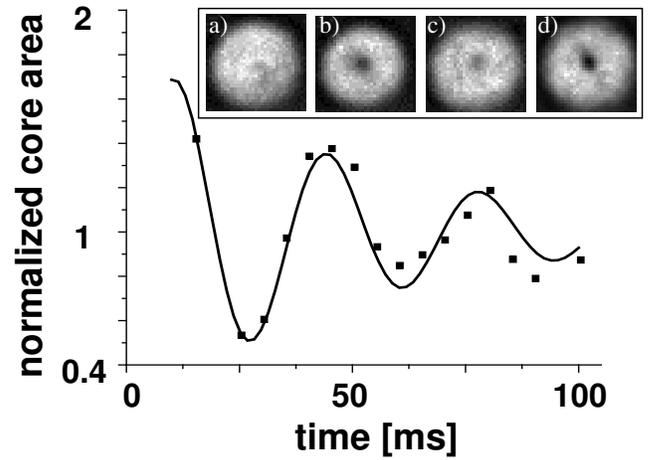,width=1\linewidth,clip=}
\end{center}
\caption {Oscillation of core area after an 8~pW, 5~ms laser pulse. Starting conditions are
$2.5 \times 10^{6}$ atoms with rotation rate
$0.9\,\omega_{\rho}$.Time given is in-trap evolution time after
end of pulse. Core area is normalized to mean of all data
points. Inset: Initial conditions $3.5 \times 10^{6}$ atoms with
rotation rate $0.78\,\omega_{\rho}$; in-trap images taken after
a 14~pW, 5~ms laser pulse followed by evolution time of (a)
20~ms, (b) 40~ms, (c) 60~ms, and (d) 80~ms. }
\label{coreosciplot}
\end{figure}

When applying stronger laser pulses, we detect clear, damped
oscillations of the core area as shown in
Fig.~\ref{coreosciplot}. We study these oscillations by
analyzing in-trap images taken after a variable evolution time
between the end of the atom-removal pulse and the expansion,
such as the ones shown in the inset of Fig.~\ref{coreosciplot}.
The oscillation frequency depends on the initial conditions; for
the case of Fig.~\ref{coreosciplot} with initially
$2.2\times10^{6}$ atoms at a rotation rate of
$0.9\,\omega_{\rho}$  we obtain a frequency of
$3.5\,\omega_{\rho}$. Decreasing the initial rotation rate of
the condensate leads to faster core-oscillation frequencies and
increases the amplitude of this oscillation; for rotation rates
below $0.9\,\omega_{\rho}$ in-trap images even show a near
complete closure of the core, followed by the core opening again
(see inset of Fig.~\ref{coreosciplot}). Presumably these
oscillations arise when a sudden removal of atoms leaves forces
from the density gradient and Coriolis forces initially out of
equilibrium. The core oscillation is not observed to be related
to an overall breathing mode of the condensate.

\begin{figure}
\begin{center}
\psfig{figure=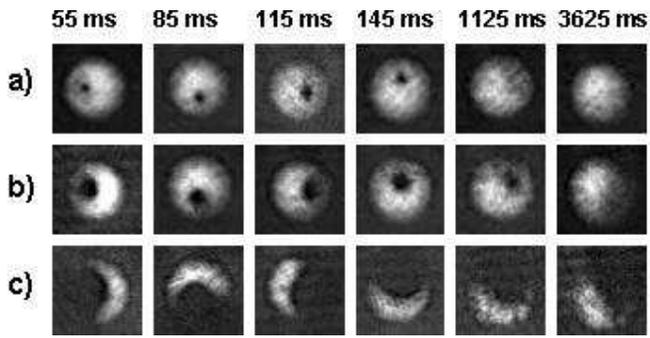,width=1\linewidth,clip=}
\end{center}
\caption {Giant core precession. Cores created by a 10~ms off-centered laser pulse with a
power of (a) 4.2~pW, (b) 33~pW, (c) 470~pW. Time given in figure
is in-trap evolution time after end of laser pulse. For
reference, a naive expectation for the lifetime of a giant
vortex given by the radius of the core divided by the speed of
sound in the surrounding cloud would be 13~ms and 30~ms for (a)
and (b), respectively. }
\label{precessions}
\end{figure}

In addition we can observe the precession
\cite{Vortexprecession} of an off-center giant vortex. For this,
the atom removal laser is deliberately offset from the center of
cloud rotation. The duration of the removal pulse is kept short
(10~ms) in comparison to the initial vortex lattice rotation
period ($2 \pi /(0.95\,\omega_{\rho}) = 126~ms$) so as to create
only a local hole. By using different laser powers, we can vary
the size of the core from a small hole as in
Fig.~\ref{precessions}(a) all the way to the extreme case of a
big hole that only leaves a crescent segment of the BEC as shown
in Fig.~\ref{precessions}(c), and the precession of these holes
can be monitored by applying variable evolution times of the
trapped BEC after the atom removal pulse is finished. In all
three cases the measured precession frequency is approximately
$\omega_{\rho}$. In the case of Fig.~\ref{precessions}(b) and
(c) we are able to follow the precession for more than 20
cycles. If, instead, the laser is left on for approximately a
full lattice rotation cycle, a complete ring can be cut out of
the condensate [Fig.~\ref{cutring}(a)], which eventually breaks
up into many individual blobs [Fig.~\ref{cutring}(c)], each of
which supports remnants of the original lattice. Both the
long-lived core precession as well as the ring structure  in
Fig.~\ref{cutring} are impressive demonstrations of the
stability of density features due to Coriolis forces in rapidly
rotating condensates.

Finally, we also observe that the sudden extraction of atoms can
launch transverse modes in the lattice \cite{Anglin}, which will
be the subject of a future publication.

\par
The work presented in this paper was funded by NSF and NIST. PE
acknowledges support by the Alexander von Humboldt Foundation.


%

\newpage

\begin{thebibliography}{10}

\bibitem[*]{qpdNIST}
Quantum Physics Division, National Institute of Standards and
Technology.

\bibitem{firstvortex}
M.~R. Matthews {\it et~al.}, Phys. Rev. Lett. {\bf 83},  2498
(1999)

\bibitem{Vortexprecession}
B.~P.~Anderson, P.~C. Haljan, C.~E. Wieman, and E.~A. Cornell,
Phys. Rev. Lett. {\bf  85}, 2857 (2000).

\bibitem{Madison2000a}
K.~W. Madison, F. Chevy, W. Wohlleben, and J. Dalibard, Phys.
Rev. Lett. {\bf 84}, 806 (2000).

\bibitem{Hodby2001a}
E. Hodby, G. Hechenblaikner, S.~A. Hopkins, O.~M. Marag{\'o},
and C.~J. Foot, Phys. Rev. Lett. {\bf 88}, 010405 (2001).

\bibitem{vortexrings}
B.~P. Anderson {\it et~al.}, Phys. Rev. Lett. {\bf 86}, 2926
(2001).

\bibitem{HauScience}
Z. Dutton, M. Budde, C. Slowe, and L.~V. Hau, Science {\bf 293},
663 (2001).

\bibitem{Madison2001a}
K.~W. Madison, F. Chevy, V. Bretin, and J. Dalibard, Phys. Rev.
Lett. {\bf 86}, 4443 (2001).

\bibitem{Abo-Shaeer2001a}
J.~R. Abo-Shaeer, C. Raman, J.~M. Vogels, and W. Ketterle,
Science {\bf 292}, 476 (2001).

\bibitem{Jilanucleation}
P.~C. Haljan, I. Coddington, P. Engels, and E.~A. Cornell, Phys.
Rev. Lett. {\bf 87},  210403 (2001).

\bibitem{GiantvortexLundh}
E. Lundh, Phys. Rev. A {\bf 65}, 043604 (2002).

\bibitem{GiantvortexUeda}
K. Kasamatsu, M. Tsubota, and M. Ueda, Phys. Rev. A {\bf 66},
053606 (2002).

\bibitem{GiantvortexBaym}
U.~R. Fischer, and G. Baym, cond-mat/0111443; G.~M. Kovoulakis,
and G. Baym, cond-mat/0212596.

\bibitem{QuarticpotentialFetter}
A.~L. Fetter, Phys. Rev. A {\bf 64}, 063608 (2001).

\bibitem{GiantvortexSalomaa}
T.~P. Simula, S.~M.~M. Virtanen, and M.~M. Salomaa, Phys. Rev. A
{\bf 65}, 033614 (2002).

\bibitem{Jilastripes}
P. Engels, I. Coddington, P.~C. Haljan, and E.~A. Cornell, Phys.
Rev. Lett. {\bf 89}, 100403 (2002).

\bibitem{Heather}
H.~J. Lewandowski, D.~M. Harber, D.~L. Whitaker, and E.~A.
Cornell, to be published.

\bibitem{MacDonald2002}
We hope eventually to use the strength of Fourier peaks of a
vortex lattice as a measure of condensation [J. Sinova, C.~B.
Hanna, and A.~H. MacDonald, cond-mat/0209374]; enhanced
expansion is needed, or the outer rings of peaks in the Fourier
plane will be lost because of insufficient imaging resolution.

\bibitem{KetterleCrystallization}
J.~R. Abo-Shaeer, C. Raman, and W. Ketterle, Phys. Rev. Lett.
{\bf 88}, 070409 (2002).

\bibitem{Feder}
D.~L. Feder and C.~W. Clark, Phys. Rev. Lett. {\bf  87}, 190401
(2001).

\bibitem{Anglin}
J.~R. Anglin and M. Crescimanno, cond-mat/0210063.

\end{thebibliography}
\end{document}